# A PHOTOMETRIC CATALOG OF 77 NEWLY-RECOGNIZED STAR CLUSTERS IN M31

PAUL HODGE[1], O. KARL KRIENKE[2], LUCIANA BIANCHI[3], PHILIP MASSEY[4] AND KNUT OLSEN[5]

[1]Astronomy Department, University of Washington, Seattle, WA 98195-1850, USA
[2]Seattle Pacific University, Seattle, WA 98119, USA
[3]Department of Physics and Astronomy, Johns Hopkins University, 3400 North Charles Street, Baltimore, MD 21218, USA
[4]Lowell Observatory, 1400 West Mars Hill Road, Flagstaff, AZ 86001, USA
[5]National Optical Astronomy Observatory, 950 North Cherry Avenue, Tucson, AZ 85719, USA

## ABSTRACT

This paper describes the results of an HST WFPC2 search for star clusters in active star-formation regions of M31. Nine of the clusters were previously cataloged and 77 are new. Our 23 fields cover key areas of the galaxy's recent star formation activity. We provide a catalog of positions and integrated magnitudes in four colors, taken with the 336W, 439W, 555W and 814W filters with the Hubble Space Telescope. A future paper will discuss the results of stellar photometry in some of the clusters in six colors, including two additional uv colors (Bianchi et al. 2010). The integrated magnitudes and colors of the clusters show a range of characteristics, but the mean color is bluer than for previous surveys, reflecting the concentration of our sample on active star forming regions. Absolute magnitudes range from $M_{555}$ = -10.3 to – 3.5. The observed luminosity function shows a nearly Gaussian distribution with a peak value at
$M_{555}$ =  -5.4 and a shoulder of unusually-bright clusters. We look in detail at two of these unusually bright examples, cluster 45 (C410) and cluster 10 (BH05). C410 lies at the core of a bright HII region. Its absolute magnitude is $M_{555}$ = -10.3. BH05 is a similar object, with an absolute magnitude of $M_{555}$ = -8.9. These two clusters are among the most luminous young clusters in M31.

Key words: galaxies:star clusters, galaxies:individual (M31, NGC 224), Local Group

## 1. INTRODUCTION

This paper is part of HST Program GO 11079 (P.I.: L. Bianchi), which was designed to provide 6 color data, including 3 ultraviolet filters, for active star-forming regions in Local Group galaxies (Bianchi et al. 2010). In this paper we report on the discovery and photometry of 77 newly-recognized star clusters in the fields obtained of M31 (NGC224), together with an additional 9 clusters that were previously recognized. We provide cluster positions and measurements of their integrated magnitudes and colors in the HST equivalent of U, B, V, and I.

## 2. OBSERVATIONS

The observations used here were obtained with the *Hubble Space Telescopes'* Wide Field /Planetary Camera 2 (WFPC2). Table 1 lists the 23 positions of the pointings in the disk of M31 used for this paper. The targets were chosen to provide data for a representative sample of M31's active star-forming regions and were centered on 20 of van den Bergh's (1964) OB associations. Their positions are plotted in Figure 1. Taking into account a small amount of overlap, the total area searched was 128 (arcmin)$^2$, which is approximately 3.0% of the active main disk, as defined by a semi-major axis of 90 arcmin and an axial ratio of 0.2.

TABLE 1. POSITIONS OF THE HST POINTINGS

| Name | RA | | | Dec | | |
| --- | --- | --- | --- | --- | --- | --- |
| | J2000 | | | J2000 | | |
| | hrs. | min. | sec. | deg. | min. | sec. |
| OB10 | 00 | 44 | 10.70 | 41 | 33 | 10.2 |
| OB22 | 00 | 41 | 29.19 | 40 | 51 | 04.3 |
| OB33 | 00 | 44 | 24.26 | 41 | 20 | 49.4 |
| OB39 | 00 | 44 | 35.70 | 41 | 25 | 05.3 |
| OB40 | 00 | 44 | 42.99 | 41 | 26 | 31.3 |
| OB41 | 00 | 44 | 50.10 | 41 | 29 | 06.8 |
| OB42 | 00 | 44 | 57.09 | 41 | 31 | 05.3 |
| OB48N | 00 | 45 | 16.75 | 41 | 39 | 30.0 |
| OB48S | 00 | 45 | 10.45 | 41 | 37 | 18.0 |
| OB51 | 00 | 45 | 42.60 | 41 | 55 | 36.3 |
| OB54 | 00 | 44 | 32.78 | 41 | 52 | 16.4 |
| OB59 | 00 | 42 | 59.66 | 41 | 37 | 27.7 |
| OB66N | 00 | 41 | 28.69 | 41 | 13 | 01.3 |
| OB66S | 00 | 41 | 24.86 | 41 | 12 | 05.6 |
| OB69 | 00 | 40 | 57.06 | 41 | 03 | 15.4 |
| OB78N | 00 | 40 | 30.87 | 40 | 45 | 15.6 |
| OB78S | 00 | 40 | 30.87 | 40 | 43 | 36.6 |
| OB99 | 00 | 46 | 31.35 | 41 | 59 | 22.8 |
| 0B102 | 00 | 46 | 33.27 | 42 | 11 | 57.1 |
| OB136 | 00 | 39 | 18.34 | 40 | 22 | 10.4 |
| OB137 | 00 | 40 | 22.96 | 40 | 52 | 40.8 |
| OB139 | 00 | 39 | 43.66 | 40 | 20 | 47.7 |
| OB157 | 00 | 47 | 01.53 | 42 | 27 | 59.6 |
| OB184 | 00 | 37 | 32.77 | 40 | 00 | 44.3 |

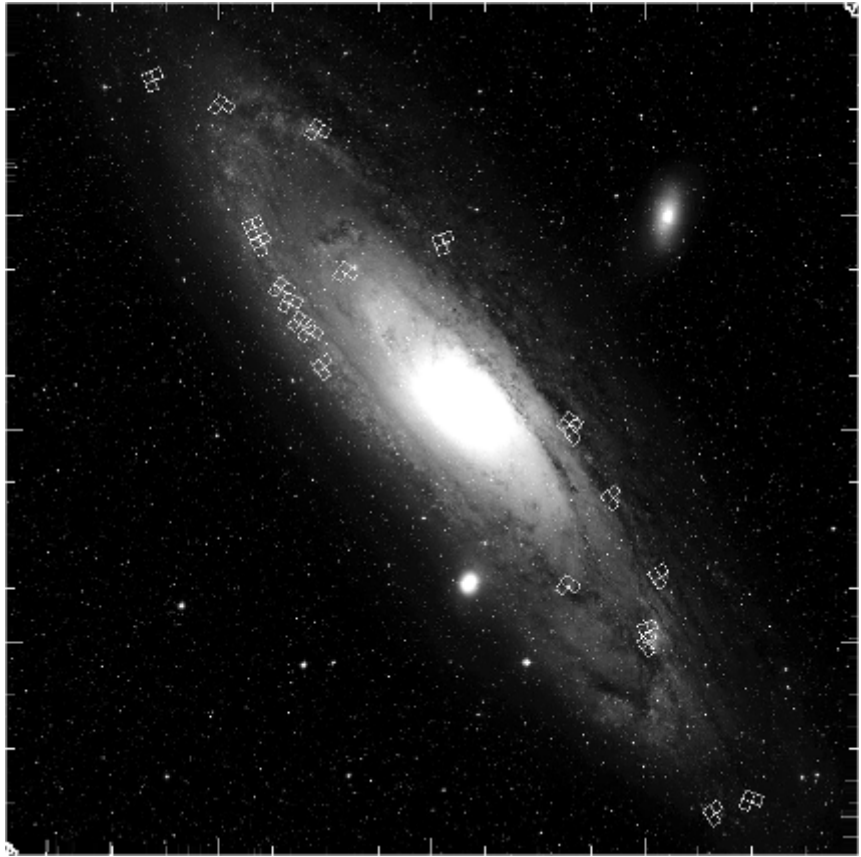

Figure 1. The location of the HST pointings used for this paper, plotted on a blue (B) photograph of M31 taken by one of the authors using the Palomar Observatory 1.2m Schmidt Telescope in 1960. Three outer pointings are beyond the limits of this image, two in the north and one in the south (see Table 1).

At each position 12 exposures were obtained, two for each filter. Filters and exposure times are listed in Table 2. Because the primary purpose of the survey was to achieve uniform S/N across the wavelength coverage (far-UV to I) for the most massive, hottest stars in regions of interest, the exposures for the three longer-wavelength filters were short. Consequently, the limiting stellar magnitude at these wavelengths is bright, averaging ~ 23.7 with the 555W filter.

TABLE 2
Filters and Exposure Times used for this Catalog

| Filter | Exposure Time (s) |
|--------|-------------------|
| F336W  | 2x300             |
| F439W  | 230+260           |
| F555W  | 2x50              |
| F814W  | 2x60              |

note: in a few cases exposures failed and were then repeated

# 3. CLUSTER IDENTIFICATION

The images in all colors were examined independently by two of us to identify conspicuous clusters. The selected clusters were limited to objects that showed a significant overdensity of stars and an underlying enhanced surface brightness, presumably made up of an unresolved cluster population. The clusters are similar to open clusters in the Milky Way and the more luminous resemble the populous clusters of the Magellanic Clouds. We noted but do not include candidates that did not have a significant number of resolved stars or a tight enough structure, using criteria based on a variety of artificial cluster experiments (described in Krienke and Hodge, 2007). Sample cluster images are given in Figure 2.

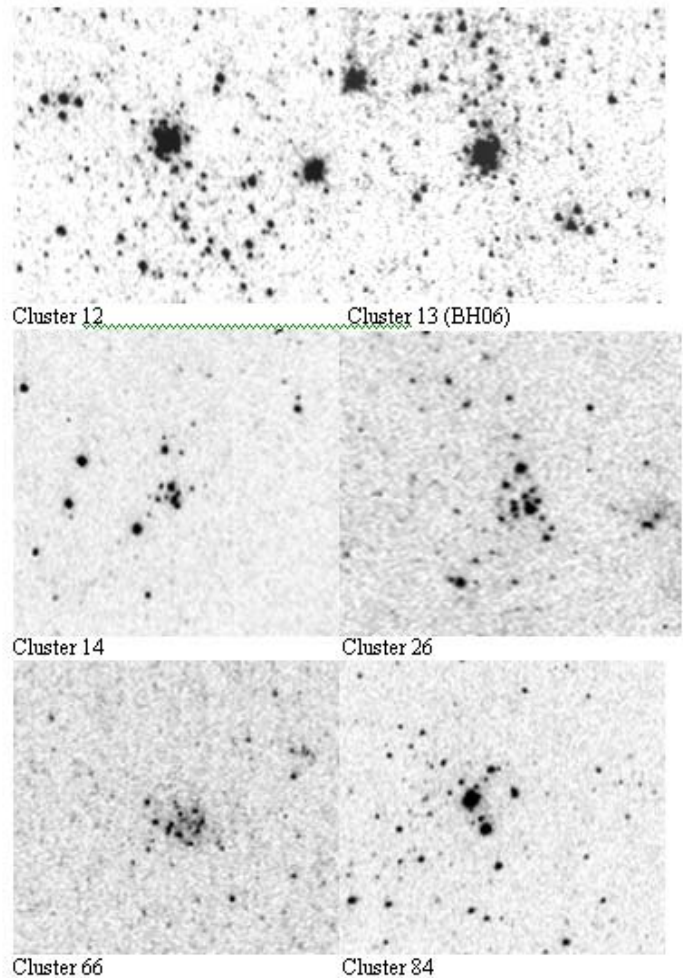

Figure 2. Six examples of clusters in the sample. The images are from the F555 exposures and the width of all images is ~14 arcsec.

The clusters were first identified as candidates on the F555 images and were then examined on images taken with the other filters. In general the clusters were most conspicuous on the F555 and F450 images. On some of the F814 images clusters tended to be almost lost in the sea of red stars of the M31 disk.

## 4. THE CLUSTER CATALOG

Table 3 is a catalog of the 86 clusters selected. We list their positions and give photometric data for the four filters, as described in Section 5. Positions were determined from the F555 images and have accuracies of ~0.2 arcsec, depending on the symmetry and structure of the cluster.

TABLE 3. INTEGRATED MAGNITUDES OF THE CLUSTERS

| no. | 336 | 336err | 439 | 439err | 555 | 555err | 814 | 814err | RA | Dec | Diam. (px) | Notes |
|---|---|---|---|---|---|---|---|---|---|---|---|---|
| 1 | 18.22 | 0.223 | 19.47 | 0.21 | 19.50 | 0.09 | 19.53 | 0.13 | 9.39204 | 40.0092 | 25 | |
| 2 | 16.41 | 0.125 | 17.52 | 0.10 | 17.23 | 0.05 | 16.98 | 0.10 | 9.93651 | 40.3493 | 49 | |
| 3 | | | 20.53 | 0.31 | 20.15 | 0.12 | 19.08 | 0.14 | 10.109 | 40.7588 | 18 | |
| 4 | 18.85 | 0.276 | 18.99 | 0.16 | 18.84 | 0.07 | 18.31 | 0.10 | 10.1136 | 40.7566 | 18 | * |
| 5 | 16.86 | 0.293 | 18.20 | 0.46 | 18.36 | 0.06 | 18.34 | 0.09 | 10.1142 | 40.7343 | 36 | |
| 6 | | | 19.64 | 0.19 | 19.29 | 0.08 | 18.41 | 0.09 | 10.1163 | 40.7626 | 31 | |
| 7 | | | 21.86 | 0.33 | 20.81 | 0.11 | 19.53 | 0.14 | 10.1234 | 40.771 | 21 | |
| 8 | 17.77 | 0.301 | 18.64 | 0.22 | 18.42 | 0.08 | 17.24 | 0.06 | 10.1239 | 40.7331 | 20 | |
| 9 | 19.23 | 0.411 | 20.10 | 0.29 | 20.05 | 0.15 | 20.31 | 0.17 | 10.1262 | 40.7256 | 16 | |
| 10 | 15.11 | 0.058 | 15.86 | 0.05 | 15.82 | 0.02 | 15.47 | 0.04 | 10.127 | 40.7581 | 54 | BH05 |
| 11 | 17.03 | 0.172 | 17.54 | 0.11 | 17.46 | 0.05 | 17.22 | 0.07 | 10.1272 | 40.7603 | 26 | |
| 12 | 16.58 | 0.152 | 17.57 | 0.11 | 17.61 | 0.07 | 17.01 | 0.05 | 10.1276 | 40.7482 | 31 | |
| 13 | 16.54 | 0.165 | 17.38 | 0.12 | 17.49 | 0.06 | 16.90 | 0.07 | 10.1277 | 40.7484 | 34 | BH06 |
| 14 | 17.65 | 0.418 | 18.84 | 0.28 | 18.91 | 0.13 | 19.05 | 0.13 | 10.1289 | 40.7126 | 19 | |
| 15 | 18.23 | 0.442 | 19.99 | 0.80 | 19.89 | 0.20 | 20.25 | 0.35 | 10.129 | 40.7262 | 19 | |
| 16 | | | 20.32 | 0.22 | 19.70 | 0.08 | 18.85 | 0.10 | 10.1299 | 40.769 | 22 | |
| 17 | 18.57 | 0.861 | 19.80 | 0.71 | 19.83 | 0.51 | 19.54 | 0.80 | 10.1306 | 40.7522 | 17 | |
| 18 | 18.04 | 0.238 | 18.94 | 0.23 | 19.08 | 0.08 | 19.05 | 0.11 | 10.1334 | 40.7449 | 18 | |
| 19 | 16.52 | 0.125 | 17.38 | 0.09 | 17.24 | 0.05 | 16.68 | 0.05 | 10.1337 | 40.7542 | 21 | |
| 20 | 17.87 | 0.218 | 18.66 | 0.15 | 18.45 | 0.06 | 17.30 | 0.05 | 10.1338 | 40.7561 | 29 | |
| 21 | 17.28 | 0.154 | 18.60 | 0.38 | 18.67 | 0.20 | 18.73 | 0.19 | 10.1396 | 40.7211 | 23 | |
| 22 | 16.67 | 0.355 | 17.50 | 0.12 | 16.85 | 0.04 | 15.95 | 0.04 | 10.1421 | 40.7218 | 20 | |
| 23 | | | 20.90 | 0.42 | 20.49 | 0.13 | 19.80 | 0.13 | 10.1456 | 40.7533 | 21 | |
| 24 | 0.48 | | 20.28 | 0.37 | 19.83 | 0.11 | 19.35 | 0.15 | 10.228 | 41.0445 | 24 | * |
| 25 | | | 20.96 | 0.31 | 20.41 | 0.11 | 19.33 | 0.12 | 10.2392 | 41.0673 | 18 | |
| 26 | 17.7 | 0.188 | 18.86 | 0.16 | 18.73 | 0.08 | 18.52 | 0.11 | 10.2435 | 41.057 | 31 | V212 |
| 27 | | | 20.20 | 0.25 | 20.06 | 0.10 | 19.33 | 0.12 | 10.3373 | 41.2001 | 31 | |
| 28 | | | 20.21 | 0.26 | 19.62 | 0.10 | 18.43 | 0.08 | 10.3446 | 41.226 | 29 | |
| 29 | | | 20.88 | 0.30 | 20.42 | 0.13 | 19.04 | 0.12 | 10.3499 | 41.1975 | 22 | |
| 30 | | | 21.55 | 0.36 | 21.30 | 0.13 | 20.33 | 0.14 | 10.3572 | 41.2085 | 17 | |
| 31 | | | 20.09 | 0.23 | 19.72 | 0.09 | 18.86 | 0.12 | 10.3594 | 41.2137 | 36 | |

| | | | | | | | | | | | |
|---|---|---|---|---|---|---|---|---|---|---|---|
| 32 | 18.19 | 0.23 | 19.24 | 0.17 | 18.80 | 0.07 | 16.62 | 0.04 | 10.3689 | 40.8505 | 26 | |
| 33 | 18.26 | 0.223 | 19.09 | 0.19 | 18.69 | 0.07 | 18.05 | 0.10 | 10.3717 | 41.2212 | 19 | |
| 34 | 17.03 | 0.191 | 17.97 | 0.11 | 17.58 | 0.07 | 16.93 | 0.05 | 10.3734 | 40.8503 | 26 | * |
| 35 | 16.84 | 0.15 | 17.85 | 0.13 | 17.56 | 0.05 | 17.18 | 0.05 | 10.3736 | 40.8495 | 43 | |
| 36 | | | 20.98 | 0.33 | 20.73 | 0.19 | 20.07 | 0.24 | 10.377 | 41.2295 | 21 | |
| 37 | | | 19.09 | 0.33 | 18.66 | 0.11 | 17.99 | 0.12 | 10.7383 | 41.623 | 48 | |
| 38 | | | 20.32 | 0.36 | 19.45 | 0.11 | 18.17 | 0.07 | 10.7441 | 41.6021 | 25 | |
| 39 | 19.84 | 0.632 | 20.24 | 0.38 | 19.55 | 0.12 | 18.00 | 0.10 | 10.7442 | 41.6149 | 32 | |
| 40 | 18.31 | 0.331 | 18.91 | 0.21 | 18.33 | 0.06 | 17.43 | 0.09 | 10.7581 | 41.6299 | 33 | |
| 41 | 20.11 | 0.468 | 20.12 | 0.24 | 19.23 | 0.08 | 18.10 | 0.08 | 10.771 | 41.6218 | 21 | |
| 42 | | | 20.88 | 0.26 | 20.39 | 0.09 | 19.64 | 0.11 | 11.056 | 41.5667 | 21 | |
| 43 | | | 21.17 | 0.28 | 20.80 | 0.09 | 19.99 | 0.11 | 11.0588 | 41.5554 | 27 | |
| 44 | | | 20.04 | 0.25 | 19.85 | 0.08 | 19.63 | 0.13 | 11.0895 | 41.3635 | 37 | |
| 45 | 15.0 | 0.064 | 16.00 | 0.05 | 15.70 | 0.02 | 15.21 | 0.03 | 11.1047 | 41.347 | 42 | C410 |
| 46 | 17.0 | 0.208 | 17.83 | 0.15 | 17.44 | 0.06 | 16.84 | 0.07 | 11.1047 | 41.3458 | 32 | |
| 47 | 19.31 | 0.395 | 19.87 | 0.29 | 19.35 | 0.11 | 17.57 | 0.07 | 11.1125 | 41.8676 | 27 | |
| 48 | 18.83 | 0.308 | 20.00 | 0.24 | 19.55 | 0.09 | 17.53 | 0.07 | 11.1146 | 41.872 | 18 | |
| 49 | 19.07 | 0.337 | 19.78 | 0.25 | 19.54 | 0.11 | 19.12 | 0.15 | 11.1211 | 41.8777 | 21 | |
| 50 | 21.2 | 2.288 | 21.44 | 1.06 | 20.76 | 0.34 | 19.72 | 0.24 | 11.1241 | 41.8654 | 18 | |
| 51 | 20.41 | 1.031 | 21.26 | 0.68 | 20.44 | 0.26 | 19.28 | 0.65 | 11.1255 | 41.8662 | 20 | |
| 52 | | | 21.61 | 0.85 | 20.92 | 0.47 | 20.15 | 0.80 | 11.1289 | 41.8821 | 18 | |
| 53 | | | 20.89 | 0.31 | 20.00 | 0.09 | 18.83 | 0.10 | 11.1323 | 41.4198 | 33 | |
| 54 | 18.02 | 0.198 | 18.35 | 0.15 | 17.83 | 0.06 | 16.89 | 0.06 | 11.1324 | 41.8802 | 27 | |
| 55 | | | 19.86 | 0.34 | 19.38 | 0.10 | 18.68 | 0.15 | 11.1353 | 41.862 | 42 | |
| 56 | 18.09 | 0.209 | 18.44 | 0.14 | 17.86 | 0.05 | 16.59 | 0.05 | 11.1382 | 41.8865 | 27 | |
| 57 | 20.5 | 1.069 | 20.81 | 0.50 | 20.78 | 0.17 | 20.07 | 0.17 | 11.1395 | 41.4233 | 18 | |
| 58 | | | 21.24 | 0.59 | 21.18 | 0.21 | 20.51 | 0.22 | 11.1401 | 41.4241 | 19 | |
| 59 | | | 19.58 | 0.46 | 19.10 | 0.14 | 18.32 | 0.18 | 11.1406 | 41.882 | 29 | |
| 60 | 20.22 | 0.534 | 20.89 | 0.32 | 20.57 | 0.13 | 19.97 | 0.16 | 11.1432 | 41.868 | 20 | |
| 61 | 17.79 | 0.185 | 18.47 | 0.14 | 18.30 | 0.06 | 17.78 | 0.07 | 11.1448 | 41.8912 | 32 | DAO69 |
| 62 | 19.37 | 0.45 | 19.96 | 0.35 | 19.38 | 0.11 | 17.58 | 0.07 | 11.1534 | 41.8768 | 19 | |
| 63 | 16.69 | 0.148 | 16.68 | 0.07 | 16.24 | 0.02 | 15.61 | 0.04 | 11.1566 | 41.4205 | 33 | |
| 64 | 19.91 | 0.567 | 20.06 | 0.26 | 19.56 | 0.10 | 18.72 | 0.11 | 11.159 | 41.8779 | 20 | |
| 65 | 16.6 | 0.128 | 17.51 | 0.10 | 17.15 | 0.04 | 16.57 | 0.05 | 11.1601 | 41.4198 | 44 | |
| 66 | | | 18.91 | 0.16 | 18.52 | 0.07 | 17.55 | 0.08 | 11.1654 | 41.4077 | 47 | B118D |
| 67 | | | 21.19 | 0.32 | 20.83 | 0.13 | 19.79 | 0.15 | 11.1698 | 41.4127 | 20 | |
| 68 | | | 20.33 | 0.26 | 20.04 | 0.09 | 19.56 | 0.13 | 11.1745 | 41.412 | 25 | |
| 69 | 18.07 | 0.218 | 19.33 | 0.29 | 19.34 | 0.09 | 19.31 | 0.14 | 11.1791 | 41.439 | 19 | |
| 70 | | | 20.78 | 0.44 | 20.46 | 0.15 | 19.37 | 0.15 | 11.2019 | 41.4798 | 25 | KHM540 |
| 71 | 16.99 | 0.39 | 17.74 | 0.21 | 17.69 | 0.09 | 17.58 | 0.09 | 11.2137 | 41.4906 | 24 | |
| 72 | | | 20.83 | 0.42 | 19.44 | 0.10 | 17.41 | 0.06 | 11.2591 | 41.5219 | 19 | |
| 73 | | | 19.36 | 0.18 | 19.49 | 0.08 | 18.90 | 0.10 | 11.2779 | 41.641 | 31 | KHM566 |
| 74 | 17.15 | 0.455 | 18.27 | 0.21 | 18.26 | 0.09 | 18.23 | 0.13 | 11.2931 | 41.6128 | 24 | |
| 75 | 16.66 | 1.11 | 17.83 | 0.23 | 17.40 | 0.10 | 16.93 | 0.13 | 11.2993 | 41.6201 | 57 | neb |
| 76 | 17.73 | 0.308 | 18.15 | 0.13 | 17.58 | 0.05 | 16.81 | 0.05 | 11.3059 | 41.6266 | 28 | |
| 77 | 16.82 | 0.116 | 18.12 | 0.11 | 18.16 | 0.05 | 18.16 | 0.07 | 11.323 | 41.6561 | 27 | |
| 78 | | | 20.82 | 0.37 | 20.39 | 0.12 | 19.94 | 0.15 | 11.3316 | 41.6647 | 26 | |
| 79 | | | 20.29 | 0.24 | 19.77 | 0.08 | 18.90 | 0.11 | 11.3378 | 41.6487 | 21 | |
| 80 | 20.11 | 0.706 | 20.38 | 0.306 | 19.99 | 0.125 | 19.27 | 0.147 | 11.4286 | 41.9273 | 44 | |
| 81 | 19.81 | 0.446 | 20.04 | 0.19 | 19.67 | 0.08 | 19.08 | 0.10 | 11.4543 | 41.9316 | 24 | |

```
82  19.32  0.348  19.59  0.21  19.36  0.09  18.76  0.12  11.6233  41.998   21
83  18.22  0.234  19.21  0.20  19.24  0.09  19.31  0.17  11.6261  42.2105  28
84  16.67  0.108  18.02  0.10  17.99  0.05  17.89  0.07  11.637   42.2099  17
85  17.99  0.231  19.08  0.19  18.98  0.09  18.58  0.11  11.638   42.1938  45
86  18.95  0.364  20.25  0.28  20.01  0.14  19.07  0.15  11.6502  42.0007  22
```

Positions are J2000.

We compared our list with the compilations of Galleti et al. (2004) and Caldwell et al. (2009) and found that 5 clusters were previously identified there and two were included in Krienke and Hodge (2008). Table 4 lists these clusters and compares our data with previous photometry (our values were converted to Johnson B and V using the precepts of Holtzman et al. (1995)). A preponderance of negative differences is probably the result of our having attempted to determine complete isophotal magnitudes, rather than magnitudes within a certain aperture. An additional known cluster, C410, did not have previous photometry.

TABLE 4. COMPARISONS WITH PUBLISHED DATA

| Name  | B          | B         |            | V          | V         |            |
|-------|------------|-----------|------------|------------|-----------|------------|
|       | this paper | published | difference | this paper | published | difference |
| BH05  |            |           |            | 15.75      | 16.03     | -0.28      |
| BH06  | 17.38      | 18.54     | -1.16      | 17.49      | 18.14     | -0.65      |
| B118D | 18.91      | 19.58     | -0.67      | 18.52      | 19.16     | -0.64      |
| V212  |            |           |            | 18.75      | 19.14     | -0.39      |
| DA069 |            |           |            | 18.30      | 18.32     | -0.02      |
| KH540 | 20.77      | 20.55     | 0.22       | 20.46      | 20.03     | 0.43       |
| KH566 | 19.22      | 19.85     | -0.63      | 19.36      | 19.54     | -0.18      |

## 5. INTEGRATED PHOTOMETRY

The values for the integrated magnitudes of the clusters listed in Table 3 were obtained using a photometric program written in IDL and described in detail in KHI. The program determines the cluster properties within a contour chosen to include most of the light, but omitting any bright foreground stars. The critical feature of the photometry is determining the background surface brightness. Because many of the clusters have a low surface brightness, the M31 background is often a significant fraction of the measured signal. The flux of the underlying stellar population, generally much older than our clusters, is also color dependent; therefore the uncertainty in background subtraction would affect more than just the flux of the cluster. Our program measures a probable background level and determines its uncertainty by sampling several (10 to 24) similarly-dimensioned fields on the image. These data are refined by Chauvenet criteria, rejecting samples with less than 0.02 probability of belonging to the population, which eliminates the possibility of having anomalously high values caused by bright foreground stars. The average of the remaining values of the background is then flux subtracted from the total flux within the cluster contour. The correction to the magnitudes due to the background subtraction was often several tenths of a magnitude. Clearly, the background correction is an important element in this photometry and it was essential that it and its uncertainty be evaluated carefully. The photometric uncertainties provided in Figure 3 and Table 3 include that of the background, which in some cases dominates the uncertainty.

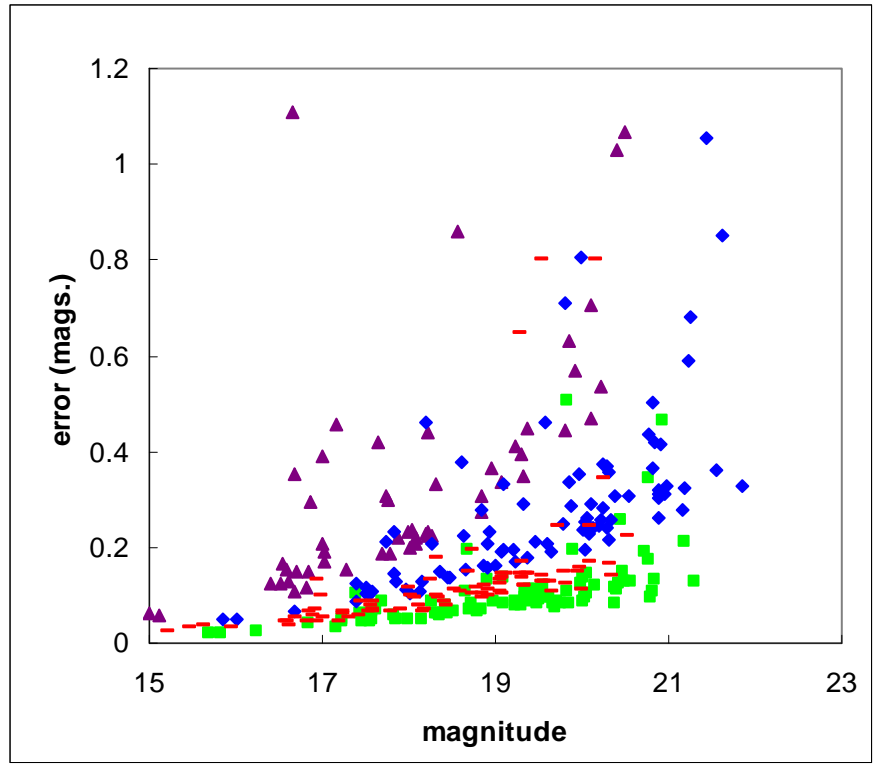

Figure 3. Photometric errors for 4 colors. Coding: F336: violet triangles, F439: blue diamonds, F555:green squares, F814: red dashes. The large F336 error at magnitude 16.66 is for cluster 75, which is involved in nebulosity.

## 6. COLOR-MAGNITUDE DIAGRAMS

It is useful to plot some of the data in traditional ways, so as to make the overall characteristics of the sample easily compared to previous studies. Therefore, we present in Figure 4 a (F439 – F555) (approximately B – V) color-magnitude diagram for the integrated measurements for the clusters, uncorrected for reddening. The distribution of points is similar to that of previous surveys of M31 clusters, except that there is a brighter lower luminosity limit, caused by the short exposures of this program, and there are relatively fewer red clusters because of the fact that the pointings are all in active star-forming regions. The mean color of the clusters in this program is F439 – F555 = 0.334, while, for comparison, the mean color for a collection of M31 clusters distributed more randomly across the disk and with longer exposures (Krienke and Hodge 2007), was B – V = 0.501.

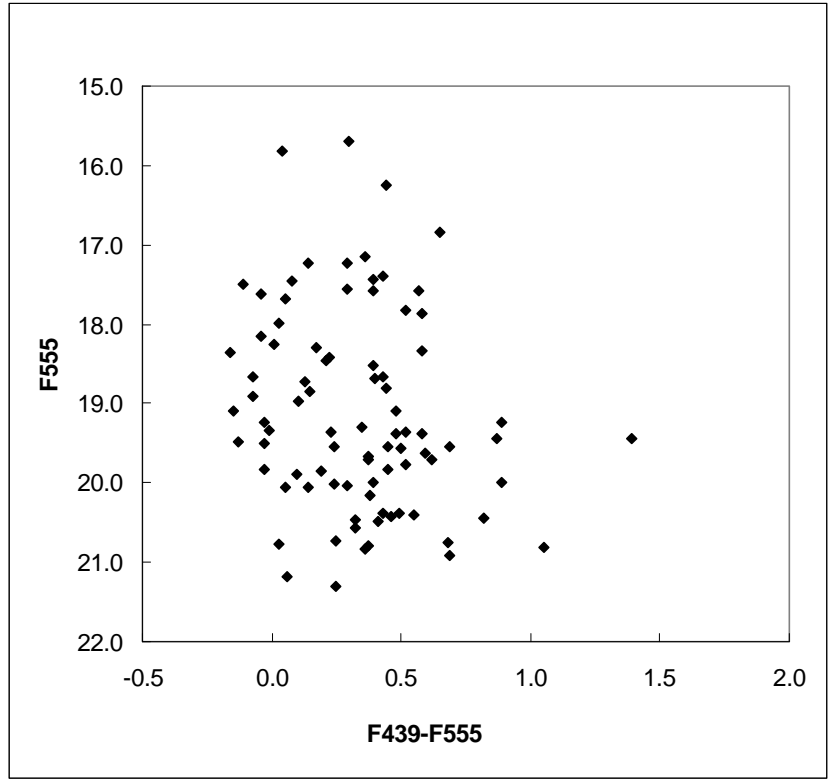

Figure 4. The F439, F555 color-magnitude diagram (closely comparable to a B,V CMD).

## 7. THE LUMINOSITY FUNCTION

The luminosity function for the current sample is shown in Figure 5, where it is compared to that for clusters in more general fields in M31, taken from Krienke and Hodge (2008). It is limited to bright clusters by the short exposures of the survey. However, it is useful for being a census of M31 star clusters that is restricted to areas of active star formation. The luminosity functions for clusters brighter than $M_V$ = ~-5.5 are very similar (the areas are approximately the same), but the deeper survey detected very many more faint clusters, as expected. Both curves suggest that there is a population of overluminous clusters, compared to the basic quasi-gaussian shape of the luminosity distribution.

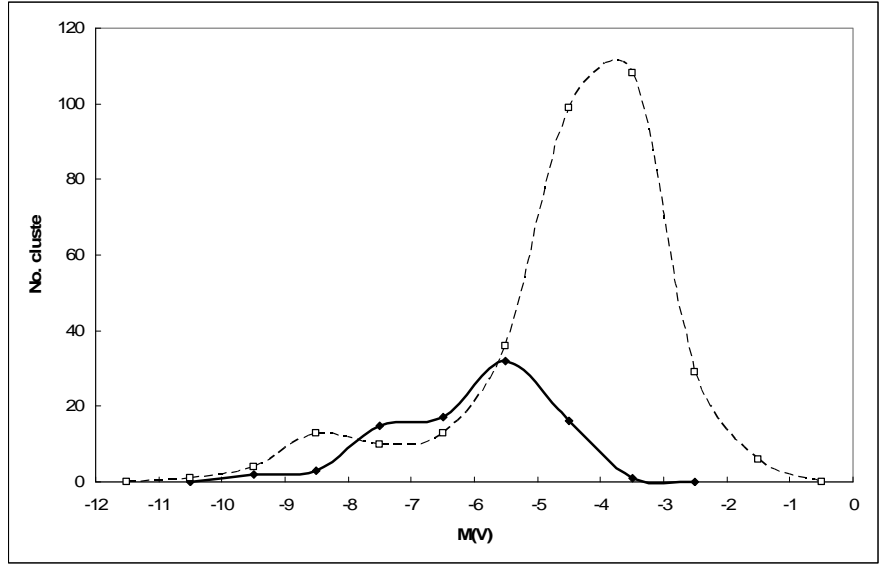

Figure 5. The luminosity functions for the present sample of clusters (solid line) and for a sample of clusters more randomly distributed across the M31 disk and observed to much fainter limits (Krienke and Hodge 2008) (dashed line).

## 8. REDDENINGS

Mean reddenings were derived by comparison of the integrated colors with Padua (Girardi et al., 2004) single age cluster models (see Bianchi et al., 2010, for further details). The mean of the different values for the reddenings is 0.28+/- 0.23 (standard deviation). This value agrees well with values found in other samples. For example, Perina et al. (2010), who determined reddenings for 25 young massive M31 clusters using HST-based color-magnitude diagrams, found an average of E(B – V) = 0.276 +/- 0.114. These values are larger than found for field early-type stars, for which Massey et al. (2007) found an average reddening of E(B – V) = 0.13 +/- 0.02.

## 9. THE GIANT CLUSTERS C410 AND BH05

Two of the most luminous clusters in the collection are interesting examples of high-luminosity, massive young clusters, of the sort recognized long ago in the Large Magellanic Cloud and increasingly being discovered in more distant galaxies (Larsen and Richtler, 1999). Fusi Pecci et al. (2005) showed that the characteristics of some 15% of the globular clusters in the Revised Bologna Catalog of Globular Clusters in M31 (Galleti et al. 2004) have photometric characteristics suggesting young ages. A recent HST color-magnitude diagram survey of 19 of these blue globular cluster candidates in M31 showed that most are massive young clusters with ages in the range 25-300 Myr (Perina et al. 2010).

Our cluster #45 was first recognized, apparently, as an unusually-bright star cluster during the preparation of the Atlas of the Andromeda Galaxy (Hodge, 1981), where it was given the designation C410. Images of it and a color-magnitude diagram are given in

Figures 6 and 7. In form it consists of a very luminous, compact cluster, surrounded by a more diffuse arrangement of luminous stars. There is a looser, small, but almost equally-bright grouping (cluster # 46) 4 arsec to the south, seen in the upper left of Figure 6.

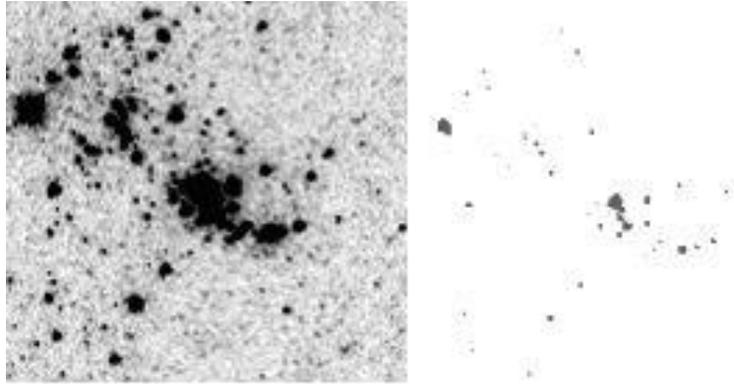

Figure 6. Images of cluster 45 (C410) (center). The left image is from a combined 100 sec exposure with the F555W filter. On the right is a shallower image at the same scale and from the same exposure, showing the inner stars. The bright central concentration is resolved further into several individual stars. Cluster 46 is above and to the left of cluster 45.

C410 has an absolute magnitude of $M_{555}$ = -10.29 and a F555 – F439 of -0.21 (both corrected for reddening). The color-magnitude diagram (figure 7) shows a well-populated luminous main sequence, indicating a very young age.

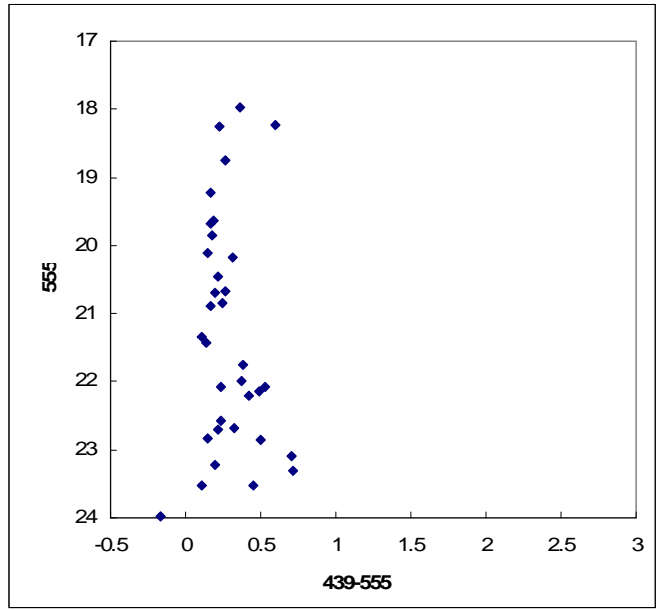

Figure 7. The color-magnitude diagram of cluster 45 (C410).

C410 and cluster 46 are near the center of one of M31's largest HII regions, Pellet 550 (Pellet et al. 1976) (Fig. 8). The large ultraviolet flux from these two objects is circumstantially responsible for the emission region. They are offset from the geometrical center of the larger Hα complex, but close to the locally most luminous emission features (Fig. 9).

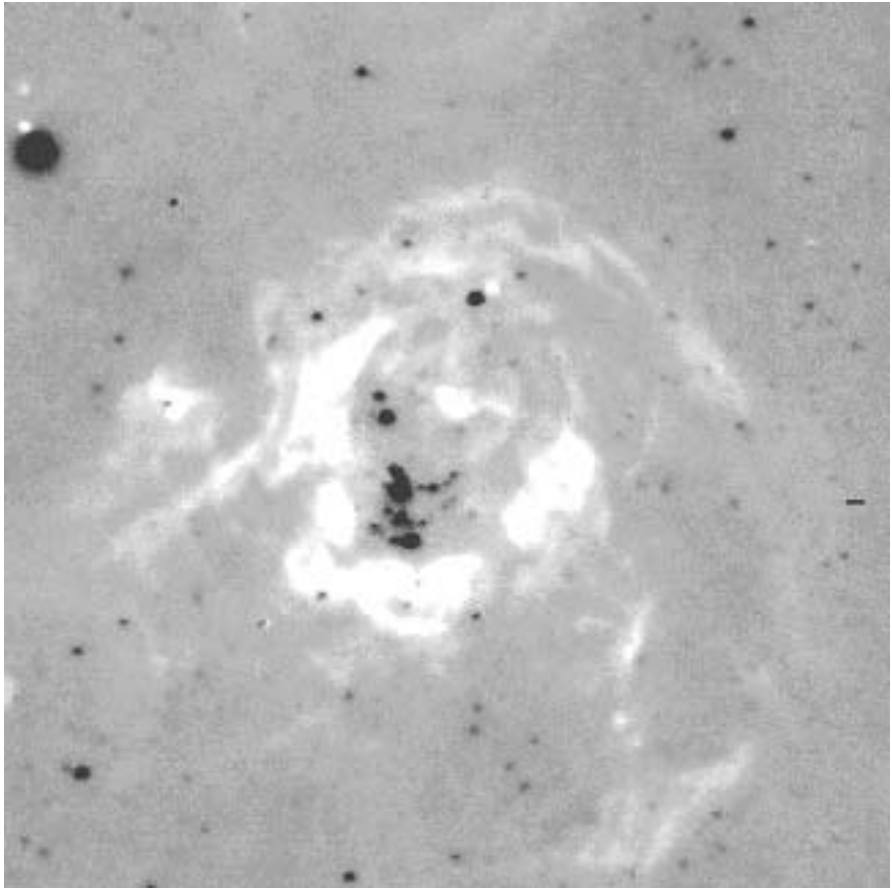

Figure 8. The HII region Pellet 550 and clusters 45 and 46. In this composite, emission is white and stellar continuum is black. C410 (cluster 45) is the brightest image at the center of the nearly linear string of stars within and to the left of the center of the HII region. This image was constructed from the images of the NOAO Local Group Galaxies Survey (Massey et al., 2008).

The cluster BH05 (Figures 9 and 10) was discovered by Barmby and Huchra (2001) who identified it as a new globular cluster. Its blue color and its color-magnitude diagram (Fig. 10) indicate that it is instead a massive young cluster. The absolute magnitude $M_{555}$ is -8.89. As its image shows, the shape is more elliptical than normal, but this is not unheard of for young massive clusters (e.g., see Perina et al. 2008, for another example in M31).

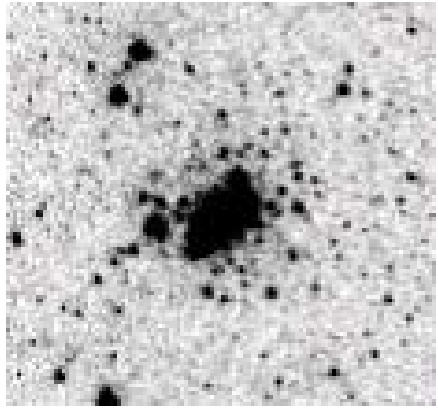

Figure 9. The cluster BH-05 (cluster 10). This is a F555 image.

The color-magnitude diagram of BH-05 (Figure 10) is similar to that of C410, with a main sequence reaching to an absolute magnitude of $M_{555}$ =-7.2. Detailed stellar photometry, including ultraviolet magnitudes, will be presented elsewhere.

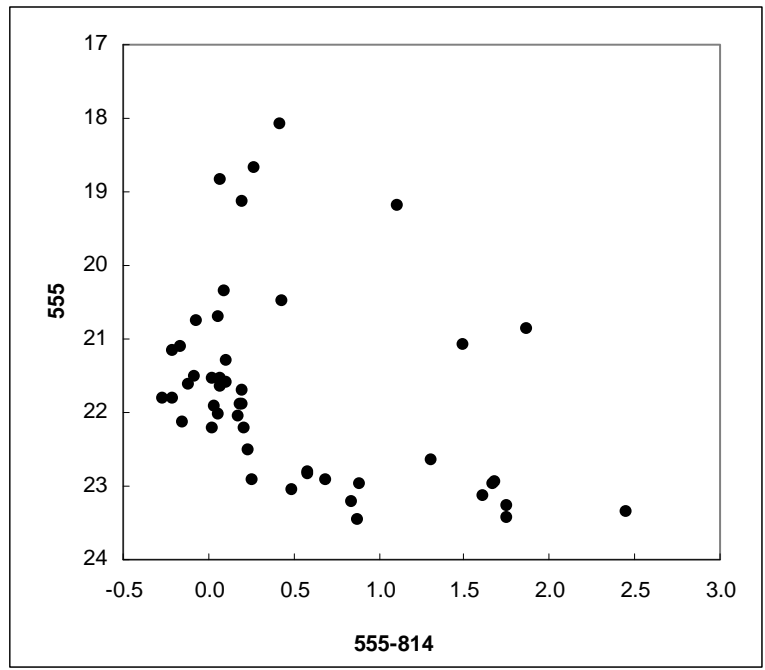

Figure 10. A color-magnitude diagram for the cluster BH05 (number 10 in Table 3).

The clusters in this survey are primarily young objects, as expected because of our selection of targets in star-forming areas if M31's disk. We postpone calculation of ages until ultraviolet photometry of them and of their individual stars is available (Bianchi 2010b), as the uv brightnesses can help to better define the stars' and clusters' spectral energy distributions.

This research was based on observations with the NASA/ESA *Hubble Space Telescope*.

Support for Program HST-GO-11079 was provided by NASA through a grant from the Space Telescope Science Institute, which is operated by the Association of Universities for Research in Astronomy, Inc., under NASA contract NAS5-26555. The color-magnitude diagrams of clusters 10 and 45 were based on HSTPHOT photometry carried out by Tanya Harrison. We are indebted to the referee, whose prompt and careful report was especially helpful.

van den Bergh, S., 1964, ApJS, 9, 65